\documentclass{article}
\usepackage{spconf,amsmath,graphicx}
\usepackage{amsfonts}  
\usepackage{amssymb}                 
\usepackage{dsfont}

\usepackage{bm}
\usepackage{algorithm}               
\usepackage{algorithmic}             
\usepackage{multirow}                

\usepackage{amssymb}                 
\usepackage{color}
\usepackage{multicol}                
\usepackage{setspace}
\usepackage{lipsum}
\usepackage{cite}

\newtheorem{lemm}{\textbf{Lemma}}
\newtheorem{Prop}{\textbf{Proposition}}

\title{Secure Transmission for IRS-Assisted MIMO mmWave Systems}
\name{Long Yang, Jiangtao Wang, Xuan Xue, Jia Shi, Yongchao Wang}
\address{State Key Laboratory of Intergrated Services Networks,
 Xidian University, Xi'an 710071 China}
\begin{document}
%
\maketitle
\newcommand\blfootnote[1]{%
\begingroup
\renewcommand\thefootnote{}\footnote{#1}%
\addtocounter{footnote}{-1}%
\endgroup
}
\begin{abstract}
In this paper, we investigate the secure beamforming design in an intelligent reflection surface (IRS) assisted millimeter wave (mmWave) system, where the hybrid beamforming (HB) and the passive beamforming (PB) are employed by the transmitter and the IRS, respectively.
To maximize the secrecy capacity, the joint optimization of HB and PB is formulated as a nonconvex problem with constant-modulus constraints. To efficiently solve such a challenging problem, the original problem is decomposed into a PB subproblem and an HB subproblem, then these two subproblems are sequentially solved by proposed algorithms.
Simulation results demonstrate the superior performance of proposed approach in comparison with the state-of-the-art works.
\blfootnote{This work was supported by the National Natural Science Foundation of China under Grants 62001361, and 61971320.}
\end{abstract}
\begin{keywords}
Millimeter wave, intelligent reflecting surface, hybrid beamforming, secure transmission
\end{keywords}
\vspace{-5pt}
\section{Introduction}
\vspace{-5pt}
Recently, hybrid beamfomring (HB) technique has demonstrated its great potentials in improving the secrecy performance of mmWave systems, where an eavesdropper (Eve) attempts to obtain the confidential information from legitimate transmitter (Alice) to legitimate receiver (Bob) \cite{ChenGLOBECOM2019,YangCL2021,WangAccess2019}.
To protect the legitimate transmission, Alice employs HB to send both the information signal (IS) and artificial noise (AN), in order to enhance the reception of Bob and suppress the reception of Eve simultaneously by utilizing the directionality and antenna gain offered by the HB.
However, such an idea cannot work properly in all possible scenarios, especially when the legitimate and eavesdropping links are highly correlated.

To overcome this issue, intelligent reflecting surface (IRS) has been introduced in the secure beamforming design of mmWave systems \cite{XiuCL2021,LuWCL2020,XiuAccess2020,QiaoWCL2020}, where the passive beamforming (PB) is employed by IRS to provide additional spatial degree-of-freedom and reconstruct wireless environment.
By integrating the full-digital beamforming (FDB) with the PB of IRS, several alternating optimization algorithms were proposed in \cite{XiuAccess2020,LuWCL2020,XiuCL2021} to maximize the secrecy capacity of mmWave systems. Since each antenna requires a dedicated radio frequency (RF) chain in FDB, the hardware complexity may be unacceptable if massive antennas are employed by Alice.
To reduce the hardware complexity of Alice, the work in \cite{QiaoWCL2020} proposed two secrecy capacity oriented alternating algorithms for downlink mmWave systems, where Bob employs only a single antenna.

In mmWave systems, the millimeter-level signal wavelength facilitates deploying multiple antennas at both Alice and Bob. Thus, it is practically meaningful to jointly design the HB and the PB for enhancing the security of IRS-assisted multiple-input-multiple-output (MIMO) mmWave systems.
However, such an issue remains unexplored so far, to the best of authors' knowledge. On the other hand, the work in \cite{QiaoWCL2020}  cannot be extended to IRS-assisted MIMO mmWave systems, since its proposed algorithms rely on the single antenna assumption for Bob.

Motivated by the above facts, we investigate the secure beamforming design for an IRS-assisted mmWave MIMO system. 
To maximize the secrecy capacity, the joint optimization of HB and PB is formulated as a non-convex problem with constant-modulus constraints. To efficiently solve such a challenging problem, the original problem is decomposed into an orthogonal forcing passive beamforming subproblem and a null-space jamming hybrid beamforming subproblem. Then these two subproblems are sequentially solved by the proposed convex approximation assisted alternating direction method of multipliers (CA-ADMM) algorithm and block coordinate descent aided orthogonal matching pursuit (BCD-OMP) algorithm. Simulation results demonstrate the superior performance of proposed algorithms.

\vspace{-5pt}
\section{System Description}\label{Sec:system}
\vspace{-5pt}
In the considered secure mmWave system, an IRS is deployed to help the legitimate transmission from Alice to Bob under the eavesdropping of Eve. The IRS consists of $N_{\rm I}$ low-cost reflecting elements, while Alice, Bob and Eve are equipped with $N_{\rm A}$, $N_{\rm B}$ and $N_{\rm E}$ antennas, respectively. Due to the use of mmWave signals, we consider that Alice employs $N_{\rm RF}(\ll N_{\rm A})$ RF chains to realize an HB architecture. To suppress the eavesdropping of Eve, Alice sends both IS ${\bf s}\in\mathcal{C}^{L_s\times 1}$ and AN ${\bf z}\in\mathcal{C}^{L_z\times 1}$ simultaneously  with using HB, where $L_{s}$ and $L_{z}$ represent the length of data streams for the IS and the AN, respectively. Therefore, the transmitted signal of Alice can be expressed as
\begin{align}\label{eqn:tran_A}
{\bf x}_{\rm A}&={\bf F}({\bf W}_{s}{\bf s}+{\bf W}_{z}{\bf z}),
\end{align}
where $\mathbb{E}[{\bf ss}^H]={\bf I}_{L_s}$, $\mathbb{E}[{\bf zz}^H]={\bf I}_{L_z}$, ${\bf F}$ is the analog beamformer, ${\bf W}_{s}$ and ${\bf W}_z$ represent the digital beamformers for the IS and the AN, respectively.

Let us denote the channel matrix between nodes $i$ and $j$ as ${\bf H}_{ij}$, where $i,j\in\{\rm{Alice(A)},\rm{Bob(B)},\rm{Eve(E)},\rm{IRS(I)}\}$ and $i\neq j$. Thus, the equivalent channel from Alice to node $i(\in\{\mathrm{B},\mathrm{E}\})$ via the IRS can be defined as
${\bf H}_{{\rm eq},i}({\bf \Theta})\triangleq{\bf H}_{{\rm A}i}+{\bf H}_{{\rm I}i}{\bf \Theta}{\bf H}_{\rm{AI}}$, where ${\bf\Theta}={\rm diag}(e^{j\theta_1},e^{j\theta_2},\cdots,e^{j\theta_{N_{\rm I}}})$ is the reflection matrix of the IRS.
Accordingly, the received signals of node $i(\in\{\mathrm{B},\mathrm{E}\})$ can be written as
{\setlength\abovedisplayskip{1.5pt}
\setlength\belowdisplayskip{1.5pt}
\begin{equation}
{\bf y}_{i}
={\bf H}_{\mathrm{eq},i}({\bf \Theta}){\bf x}_{\rm A}+{\bf n}_{i},\label{eqn:rec_B}
\end{equation}}
where ${\bf n}_{i}\sim \mathcal{CN}({\bf{0}}, \sigma^2{\bf I})$ represent the vector of zero-mean additive white Gaussian noise (AWGN) with variance $\sigma^2$.
Then, by the received signal shown in \eqref{eqn:rec_B}, the secrecy capacity can be expressed as
\begin{align}\label{eqn:secap}
&C_{\rm{sec}}({\bf W}_s, {\bf W}_z,{\bf F},{\bf \Theta})=\big\{\log{\rm det} \left[{\bf I}_{N_{\rm B}}+{\bf S}_{\rm B}({\bf \Theta}){\bf C}_{\rm B}^{-1}({\bf \Theta}) \right]\nonumber\\
&\hspace{20mm}-\log{\rm det} \left[{\bf I}_{N_{\rm E}}+{\bf S}_{\rm E}({\bf \Theta}){\bf C}_{\rm E}^{-1}({\bf \Theta}) \right]\big\}^+,
\end{align}
where ${\bf S}_i({\bf \Theta})\triangleq{\bf H}_{{\rm eq},i}({\bf \Theta}){\bf F}{\bf W}_{s}{\bf W}_{s}^H{\bf F}^H{\bf H}_{{\rm eq},i}^H({\bf \Theta})$, ${\bf C}_i({\bf \Theta})\triangleq{\bf H}_{{\rm eq},i}({\bf \Theta}){\bf F}{\bf W}_{z}{\bf W}_{z}^H{\bf F}^H{\bf H}_{{\rm eq},i}^H({\bf \Theta})+\sigma^2{\bf I}_{N_i}$, for $i(\in\{\mathrm{B},\mathrm{E}\})$.

\vspace{-10pt}
\section{Secrecy Capacity Maximization Problem}\label{Sec:problem}
\vspace{-10pt}

To maximize the secrecy capacity defined in \eqref{eqn:secap}, the joint optimization of PB and HB can be formulated as follows
{\setlength\abovedisplayskip{1.5pt}
\setlength\belowdisplayskip{1.5pt}
\begin{subequations}\label{eqn:know_problem}
\begin{align}
(\mathcal{P}_0):\hspace{1mm}\max_{{\bf W}_{s},{\bf W}_{z},{\bf F},{\bf \Theta}}&\hspace{1mm}C_{\rm{sec}}({\bf W}_{s},{\bf W}_{z},{\bf F},{\bf \Theta})
\label{eqn:objective_function}\\
{\rm s.t.}\hspace{6.5mm}
&\hspace{1mm}\|{\bf F}{\bf W}_{s}\|_{\rm F}^2+\|{\bf F}{\bf W}_{z}\|_{\rm F}^2\leq P_{\max},\label{eqn:powcons}\\
&\hspace{1mm}|{\bf F}|=\frac{1}{\sqrt{N_{\rm A}}}{\bf 1},\label{eqn:analogcon}\\
&\hspace{1mm}{\bf\Theta}={\rm diag}(e^{j\theta_1},e^{j\theta_2},\cdots,e^{j\theta_{N_{\rm I}}}),\label{eqn:IRScm1}\\
&\hspace{1mm}\theta_n\in [0, 2\pi),\hspace{2mm} n=1,\cdots,N_{\rm I}.\label{eqn:IRScm2}
\end{align}
\end{subequations}}
In the above problem, constraints \eqref{eqn:powcons}$\sim$\eqref{eqn:IRScm2} are explained as follows: \eqref{eqn:powcons} represents the transmit power constraint, where $P_{\max}$ is the maximum transmit power; \eqref{eqn:analogcon} means the constant-modulus constraint for analog beamformer at Alice; \eqref{eqn:IRScm1} and \eqref{eqn:IRScm2} represent the continuous diagonal constant-modules constraint for PB. Note that the PB and HB are highly coupled in objective function \eqref{eqn:objective_function}.
To decouple the formulated problem, we equivalently reformulate the original problem to the following subproblem with only variable ${\bf \Theta}$:
\begin{equation*}
({\mathcal P}_1):\hspace{3mm}\max_{\bf \Theta}\hspace{3mm}C_{\rm{sec}}({\bf \Theta}),\hspace{3mm}{\rm s.t.}\hspace{3mm} \eqref{eqn:IRScm1}, \eqref{eqn:IRScm2},\nonumber
\end{equation*}
where $C_{\rm{sec}}({\bf \Theta})$ is the short-hand notation for $C_{\rm{sec}}[{\bf W}^{\star}_{s}({\bf \Theta}),$ ${\bf W}^{\star}_{z}({\bf \Theta}),{\bf F}^{\star}({\bf \Theta}),{\bf \Theta},\psi^{\star}({\bf \Theta})]$, with  $\psi^{\star}({\bf \Theta})$, ${\bf F}^{\star}({\bf \Theta})$, ${\bf W}_s^{\star}({\bf \Theta})$ and ${\bf W}_z^{\star}({\bf \Theta})$ being the optimal solutions of $\psi$, ${\bf F}$, ${\bf W}_s$ and ${\bf W}_z$.

Since the exact expression of $C_{\rm{\rm sec}}({\bf \Theta})$ in subproblem ($\mathcal{P}_1$) cannot be explicitly formulated, we have the following lemma to derive an approximate problem to subproblem ($\mathcal{P}_1$).
\begin{lemm}\label{lem:lowboundCsec}
With appropriate parameter $\alpha_B$, the solution of subproblem $(\mathcal{P}_1)$ can be approximately obtained by solving the \textbf{Orthogonal-Forcing Passive Beamforming (OF-PB)} subproblem  shown as
\begin{align}\label{eqn:PBsub}
&\min_{\bf \Theta}\hspace{2mm}\|{\bf H}_{\rm{eq,B}}({\bf \Theta}){\bf H}_{\rm{eq,E}}^H({\bf \Theta})\|_{\rm F}^2-\alpha_B\|{\bf H}_{\rm{eq,B}}({\bf \Theta})\|_{\rm F}^2,\\
&{\rm s.t.}
\hspace{2mm}\eqref{eqn:IRScm1}, \eqref{eqn:IRScm2},
\end{align}
\end{lemm}
{\bf Proof:} \emph{Due to the page limit,  the detailed proof will be presented in our future paper.}

Applying the optimal solution of OF-PB subproblem, denoted by ${\bf \Theta}^\star$, into original problem $(\mathcal{P}_0)$, we further obtain the following  \textbf{Null-Space Jamming Hybrid Beamforming (NSJ-HB)} subproblem:
{\setlength\abovedisplayskip{1.5pt}
\setlength\belowdisplayskip{1.5pt}
\begin{align}
\label{eqn:p2reformu}
&\max_{{\bf W}_{s},{\bf W}_{z},{\bf F}}\hspace{3mm}C_{\rm{sec}}({\bf W}_{s},{\bf W}_{z},{\bf F},{\bf \Theta}^{\star})\hspace{3mm}{\rm s.t.}\hspace{3mm}
\eqref{eqn:powcons},\eqref{eqn:analogcon}.
\end{align}}
With the above decomposition, the suboptimal solution of original problem $({\mathcal{P}}_0)$ can be obtained by sequentially solving OF-PB subproblem and NSJ-HB subproblem by algorithms that will be proposed in the next section.
\vspace{-12pt}
\section{Proposed algorithms for OF-PB and NSJ-HB subproblems}\label{sec:algorithm}
\vspace{-10pt}
\subsection{CA-ADMM algorithm for solving OF-PB subproblem}
\vspace{-5pt}
By introducing a variable ${\bf x}\in{\mathcal{C}}^{N_{\rm I}}$, which denotes the vector of main diagonal elements of ${\bf \Theta}$, the objective function of OF-PB subproblem can be reformulated as
{\setlength\abovedisplayskip{1.5pt}
\setlength\belowdisplayskip{1.5pt}
\begin{align}\label{eqn:obj2}
&\|\tilde{\bf h}_{a}+{\bf H}_{PB,1}{\bf x}^*+{\bf H}_{PB,2}{\bf x}+ {\bf H}_{PB,3}{\rm vec}({\bf x}{\bf x}^H)\|_{\rm 2}^2\nonumber\\
&-\alpha_B\|{\bf h}_{\text {AB}}+{\bf H}_{PB,4}{\bf x}\|_2^2,
\end{align}}
where $\tilde{\bf h}_a={\rm vec}\big({\bf H}_{\rm {AB}}{\bf H}_{\rm {AE}}^H\big)$, ${\bf H}_{PB,1}={\bf H}_{\rm{IE}}^*\odot({\bf H}_{\rm {AB}}{\bf H}_{\rm{AI}}^H)$, ${\bf H}_{PB,2}=({\bf H}_{\rm{AI}}{\bf H}_{\rm {AE}}^H)^T\odot{\bf H}_{\rm{IB}}$, ${\bf H}_{PB,4}={\bf H}_{\rm{AI}}^T\odot{\bf H}_{\rm{IB}}$, ${\bf H}_{PB,3}=({\bf H}_{\rm{IE}}^*\otimes{\bf H}_{\rm{IB}}){\rm diag}\big({\rm vec}({\bf H}_{\rm{AI}}{\bf H}_{\rm{AI}}^H)\big)$ and ${\bf h}_{\rm {AB}}={\rm vec}({\bf H}_{\rm {AB}})$.

Then, to deal with the constant-modulus constraint and fourth-order objective function \eqref{eqn:obj2}, we introduce variable ${\bf y}_1$ and ${\bf y}_2$ that satisfies ${\bf y}_1={\bf y}_2={\bf x}$. Thus, the objective function of problem \eqref{eqn:obj2} is reformulated as
{\setlength\abovedisplayskip{1.5pt}
\setlength\belowdisplayskip{1.5pt}
\begin{align}
&f({\bf x},{\bf y}_1,{\bf y}_2)\\
&\triangleq\|\tilde{\bf h}_{a}+{\bf H}_{PB,1}{\bf x}^*+{\bf H}_{PB,2}{\bf x}+ {\bf H}_{PB,3}{\rm vec}({\bf y}_1{\bf y}_1^H)\|_{\rm 2}^2\nonumber\\
&-\alpha_B\|{\bf h}_{\text {AB}}+{\bf H}_{PB,4}{\bf x}\|_2^2-{\bf x}^H{\bf H}_{PB,5}{\bf x}\nonumber\\
&-2{\rm Re}({\bf x}^T{\bf H}_{PB,6}{\bf x})+{\bf y}_2^H{\bf H}_{PB,5}{\bf y}_2+2{\rm Re}({\bf y}_2^T{\bf H}_{PB,6}{\bf y}_2),\nonumber
\end{align}}
where ${\bf H}_{PB,6}={\bf H}_{PB,1}^H{\bf H}_{PB,2}$ and ${\bf H}_{PB,5}=({\bf H}_{PB,1}^H{\bf H}_{PB,1})^*$ $+{\bf H}_{PB,2}^H{\bf H}_{PB,2}-\alpha_B{\bf H}_{PB,4}^H{\bf H}_{PB,4}$.

Therefore, problem \eqref{eqn:obj2} is equivalent to be
{\setlength\abovedisplayskip{1.5pt}
\setlength\belowdisplayskip{1.5pt}
\begin{align} \label{eqn:pb2}
&\min_{|{\bf x}|={\bf 1},{\bf y}_1,{\bf y}_2}f({\bf x},{\bf y}_1,{\bf y}_2)\hspace{4mm}{\rm s.t.}\hspace{2mm} {\bf y}_1={\bf x},\hspace{2mm}{\bf y}_2={\bf x}.
\end{align}}
Then, the augmented Lagrangian function of problem \eqref{eqn:pb2} is given by
{\setlength\abovedisplayskip{1.5pt}
\setlength\belowdisplayskip{1.5pt}
\begin{align}\label{eqn:lagra}
&{\mathcal L}({\bf x},{\bf y}_1,{\bf y}_2,{\boldsymbol \lambda}_1,{\boldsymbol \lambda}_2)\nonumber\\
&=f({\bf x},{\bf y}_1,{\bf y}_2)+{\rm Re}({\boldsymbol \lambda}_1^H({\bf y}_1-{\bf x})+{\boldsymbol \lambda}_2^H({\bf y}_2-{\bf x}))\nonumber\\
&\hspace{5mm}+\frac{\rho_1}{2}\|{\bf y}_1-{\bf x}\|_2^2+\frac{\rho_2}{2}\|{\bf y}_2-{\bf x}\|_2^2
\end{align}}
Note that the constant-modulus constraint is only imposed to variable ${\bf x}$ and are not related to variables ${\bf y}_1$ and ${\bf y}_2$. Therefore, denoting $k$ as the iteration number, the ADMM algorithm framework for solving problem \eqref{eqn:lagra} is given by
\vspace{-2mm}
\begin{subequations}\label{eqn:admm}
\begin{align}
&{\bf x}^{k+1} = \underset{|{\bf x}|={\bf 1}}{{\rm arg}\min}\hspace{2mm}{\mathcal L}({\bf x},{\bf y}_1^k,{\bf y}_2^k,{\boldsymbol \lambda}_1^k,{\boldsymbol \lambda}_2^k)\label{eqn:admmx}\\
&{\bf y}_1^{k+1} = \underset{{\bf y}_1}{{\rm arg} \min} \hspace{2mm}{\mathcal L}({\bf x}^{k+1},{\bf y}_1,{\bf y}_2^k,{\boldsymbol \lambda}_1^k,{\boldsymbol \lambda}_2^k)\label{eqn:admmy}\\
&{\bf y}_2^{k+1} = \underset{{\bf y}_2}{{\rm arg} \min} \hspace{2mm}{\mathcal L}({\bf x}^{k+1},{\bf y}_1^{k+1},{\bf y}_2,{\boldsymbol \lambda}_1^k,{\boldsymbol \lambda}_2^k)\label{eqn:admmz}\\
&{\boldsymbol \lambda}_1^{k+1}={\boldsymbol \lambda}_1^{k}+\rho_1({\bf y}_1^{k+1}-{\bf x}^{k+1})\label{eqn:admmlambda1}\\
&{\boldsymbol \lambda}_2^{k+1}={\boldsymbol \lambda}_2^{k}+\rho_2({\bf y}_2^{k+1}-{\bf x}^{k+1})\label{eqn:admmlambda2}
\end{align}
\end{subequations}

As observed from the ADMM algorithm framework \eqref{eqn:admm}, the closed-form solutions of the constant-modulus constrained linear programming problem \eqref{eqn:admmx} and the non-constrained quadratic programming problem \eqref{eqn:admmz} can be derived following \cite{Liang2016TSP}. The main challenge of utilizing the ADMM algorithm lies on solving problem \eqref{eqn:admmy}.
To solve the fourth-order non-convex problem \eqref{eqn:admmy}, we first introduce a variable $\hat{\bf y}_1=[{\rm Re}({\bf y}_1);{\rm Im}({\bf y}_1)]$, and convert ${\mathcal L}({\bf x}^{k+1},{\bf y}_1,{\bf y}_2^k,{\boldsymbol \lambda}_1^k,{\boldsymbol \lambda}_2^k)$ to be an equivalent function with real-value variable as follows:
{\small\begin{equation}\label{eqn:y_Lagreal}
\hat{\mathcal{L}}_y^k(\hat{\bf y}_1)=f_y^k(\hat{\bf y}_1)+{\hat{\boldsymbol \lambda}}_1^{k^T}({\hat{\bf y}_1}-{\hat{\bf x}}^{k+1}))+\frac{\rho_1}{2}\|{\hat{\bf y}_1}-{\hat{\bf x}}^{k+1}\|_2^2,
\end{equation}}
where $f_y^k(\hat{\bf y}_1)=\|{\hat{\bf H}}_{PB,3}[({\bf K}_1{\hat{\bf y}_1\otimes{\bf K}_2}+{\bf K}_3{\hat{\bf y}_1\otimes{\bf K}_4}){\hat{\bf y}_1}]+{\hat{\bf a}}_{y}^k\|_2^2$, with $\hat{\bf H}_{PB,3}=\begin{bmatrix}{\rm Re}({\bf H}_{PB,3}), &-{\rm Im}({\bf H}_{PB,3})\\{\rm Im}({\bf H}_{PB,3}), &{\rm Re}({\bf H}_{PB,3})\end{bmatrix}, \hat{\bf a}_y^k=\begin{bmatrix}{\rm Re}({\bf a}_y^k)\\{\rm Im}({\bf a}_y^k)\end{bmatrix},$ ${\bf a}_{y}^{k}=\tilde{\bf h}_{a}+{\bf H}_{PB,1}{\bf x}^{k+1^*}+{\bf H}_{PB,2}{\bf x}^{k+1}$, ${\hat{\boldsymbol \lambda}}_1^k=\begin{bmatrix}{\rm Re}({\boldsymbol \lambda}_1^k)\\{\rm Im}({\boldsymbol \lambda}_1^k)\end{bmatrix}, {\hat{\bf x}}^{k+1}=\begin{bmatrix}{\rm Re}({\bf x}^{k+1})\\{\rm Im}({\bf x}^{k+1})\end{bmatrix}, {\bf K}_1=\begin{bmatrix}{\bf I}_N, &{\bf 0}\\{\bf 0}, &-{\bf I}_N\end{bmatrix},$ ${\bf K}_2=\begin{bmatrix}{\bf I}_N,{\bf 0}_{N\times N}\end{bmatrix}, {\bf K}_3=\begin{bmatrix}{\bf 0},&{\bf I}_N\\{\bf I}_N, &{\bf 0}\end{bmatrix},{\bf K}_4=\begin{bmatrix}{\bf 0}_{N\times N}, {\bf I}_N\end{bmatrix}.$

To solve this fourth-order problem, we can find that $\hat{\mathcal{L}}_y^k(\hat{\bf y}_1)$ is Lipschitz continuous when $L_y>8c_2^2|\sum_i\sum_j\tilde{\bf H}_{3}$ $(i,j)|+4c_2^2|\sum_i\sum_j\tilde{\bf H}_{3}(i,j)|+4\|\tilde{\bf H}_{PB,3}^T{\tilde{\bf a}}_{y}\|_1$,
with $c_2=\max\{c,1\}$, $\tilde{\bf H}_{3}=\hat{\bf H}_{PB,3}^T\hat{\bf H}_{PB,3}$, $\tilde{\bf H}_{PB,3}(i,j)=|\hat{\bf H}_{PB,3}(i,j)|$, and ${\tilde{\bf a}}_{y}(i)=|\tilde{\bf h}_{a}(i)+\sum_i{\bf H}_{PB,1}(i,:)+\sum_i{\bf H}_{PB,2}(i,:)|$. Therefore, the upper-bound quadratic function of $\hat{\mathcal{L}}_y^k(\hat{\bf y}_1)$ can be formulated as
{\setlength\abovedisplayskip{1.5pt}
\setlength\belowdisplayskip{1.5pt}
\begin{align}\label{eqn:uyk}
\mathcal{U}_y^k(\hat{\bf y}_1)\triangleq& f_y^k(\hat{\bf x}^{k+1})+(\nabla f_y^k(\hat{\bf x}^{k+1})+{\hat{\boldsymbol \lambda}}_1^{k})^T({\hat{\bf y}_1}-{\hat{\bf x}}^{k+1})\nonumber\\
&+\frac{\rho_1+L_y}{2}\|{\hat{\bf y}_1}-{\hat{\bf x}}^{k+1}\|_2^2
\end{align}}
Applying the convex approximation (CA) technique, we customize the ADMM algorithm by minimizing $\mathcal{U}_y^k(\hat{\bf y}_1)$ instead of \eqref{eqn:admmy}. The CA-ADMM algorithm is summarized in Algorithm \ref{Alg:cADMM_IRS}.
{\setlength\abovedisplayskip{1.5pt}
\setlength\belowdisplayskip{1.5pt}
\begin{algorithm}[t]
{\baselineskip 12pt
\caption{The CA-ADMM Algorithm}
\label{Alg:cADMM_IRS}
\begin{algorithmic}[1]
\STATE \textbf{Initialize}: Set Lipschitz constant $L_y$ and iteration index $k=1$, and initialize $\{{\bf x}^1, {\bf y}_1^1, {\bf y}_2^1, {\boldsymbol \lambda}_1^{1}, {\boldsymbol \lambda}_2^{1}\}$ randomly.
\WHILE {$\|{\bf x}^k-{\bf y}_1^k\|_2^2+\|{\bf x}^k-{\bf y}_2^k\|_2^2\geq \epsilon_1$}
\STATE Compute ${\bf x}^{k+1}$ via \eqref{eqn:admmx};
\STATE Compute ${\bf y}_1^{k+1}$ by minimizing $\mathcal{U}_y^k(\hat{\bf y}_1)$ in \eqref{eqn:uyk} and compute ${\bf y}_2^{k+1}$ via \eqref{eqn:admmz};
\STATE Compute ${\boldsymbol \lambda}_1^{k+1}$ and ${\boldsymbol \lambda}_2^{k+1}$ in parallel via \eqref{eqn:admmlambda1} and \eqref{eqn:admmlambda2}, respectively.
\ENDWHILE
\RETURN{${\bf \Theta}^{\star}={\rm {diag}}({\bf x}^{k+1})$.}
\end{algorithmic}
}
\end{algorithm}}

\vspace{-10pt}
\subsection{BCD-OMP algorithm for solving NSJ-HB subproblem}\label{sec:solveHP}
\vspace{-5pt}
To solve NSJ-HB subproblem \eqref{eqn:p2reformu}, we first denote IS full-digital beamforming (ISFDB) as $\tilde{\bf W}_{s}\triangleq{\bf F}{\bf W}_{s}$ and AN full-digital beamforming (ANFDB) as $\tilde{\bf W}_{z}\triangleq{\bf F}{\bf W}_{z}$. Replacing the variables of NSJ-HB subproblem \eqref{eqn:p2reformu} as the ISFDB $\tilde{\bf W}_{s}$ and the ANFDB $\tilde{\bf W}_{z}$, we will get the full-digital beamforming (FDB) problem, which can be solved by applying the Block Coordinate Descent (BCD) algorithm proposed in \cite{Pan2020TCOM}.

Second, the hybrid beamforming ${\bf W}_{s}$ and ${\bf F}$ can be solved by minimizing the Euclidean distance from the optimal ISFDB $\tilde{\bf W}_{s}^{\star}$, leading to the following problem
{\setlength\abovedisplayskip{1.5pt}
\setlength\belowdisplayskip{1.5pt}
\begin{align} \label{eqn:hpproblem}
&\min_{{\bf F},{\bf W}_{s}}\hspace{3mm} \|\tilde{\bf W}_{s}^{\star}-{\bf F}{\bf W}_{s}\|_{\rm F}^2, \hspace{6mm}{\rm s.t.}
\hspace{2mm}\eqref{eqn:analogcon}.
\end{align}}
The above problem \eqref{eqn:hpproblem} can be solved by using Orthogonal Matching Pursuit (OMP) algorithm proposed in \cite{AyachTcom2014}.

Finally, the optimal digital artificial noise beamforming ${\bf W}_{z}^{\star}$ can be obtained by letting ${\bf W}_{z}^{\star}$ lie on the null space of the effective full-digital channel ${\bf H}_{\rm{eq,B}}({\bf \Theta}^{\star}){\bf F}^{\star}$.
The BCD-OMP algorithm is summarized in \textbf{Algorithm} \ref{Alg:BCD-OMP}.
{\setlength\abovedisplayskip{1.5pt}
\setlength\belowdisplayskip{1.5pt}
\begin{algorithm}[t]
{\baselineskip 12pt
\caption{The BCD-OMP algorithm}
\label{Alg:BCD-OMP}
\begin{algorithmic}[1]
\STATE \textbf{Input}: ${\bf \Theta}^{\star}$ solved by CA-ADMM algorithm;
\STATE Compute the optimal FDB $\tilde{\bf W}_{s}^{\star}$ by applying BCD algorithm in \cite{Pan2020TCOM};
\STATE Compute the optimal digital transmit beamforming ${\bf W}_{s}^{\star}$ and the optimal analog beamforming ${\bf F}^{\star}$ with optimized FDB $\tilde{\bf W}_{s}^{\star}$ via \eqref{eqn:hpproblem} by using OMP algorithm proposed in \cite{AyachTcom2014};
\STATE Compute the optimal digital artificial noise beamforming ${\bf W}_{z}^{\star}$ by letting ${\bf W}_{z}^{\star}$ lie on the null space of the effective full-digital channel ${\bf H}_{\rm{eq,B}}({\bf \Theta}^{\star}){\bf F}^{\star}$.
\RETURN{${\bf W}_{s}^{\star}$, ${\bf F}^{\star}$, and ${\bf W}_{z}^{\star}$.}
\end{algorithmic}
}
\end{algorithm}}

\vspace{-5mm}
\section{Convergence and complexity analysis}\label{Sec:performance}
\vspace{-5mm}
\subsection{Convergence Analyses of the CA-ADMM algorithm}
We have the following proposition to show that the proposed CA-ADMM algorithm converges to a stationary point under some conditions.
\vspace{-5pt}
\begin{Prop}\label{prop:optimal}
If penalty parameter $\rho_1,\rho_2$ and Lipschitz constant $L_y$ satisfy the inequalities $\frac{\rho_1+\rho_2}{2}-\frac{8L_y^2\rho_1+32L_y^3}{\rho_1^2}\geq 0$, $\frac{\rho_1-7L_y}{2}-\frac{2\rho_1+8L_y}{\rho_1^2}\geq 0$, and $\rho_1-5L_y\geq 0$, the proposed CA-ADMM algorithm converges to a stationary point of the equivalent OF-PB subproblem \eqref{eqn:pb2}.
\end{Prop}
{\bf Proof:} \emph{Due to the page limit,  the detailed proof will be presented in our future paper.}
\vspace{-10pt}
\subsection{Complexity Analysis}
\vspace{-5pt}
The computational complexity of proposed CA-ADMM and BCD-OMP algorithms is analyzed as follows:

 \textbf{Complexity of CA-ADMM Algorithm}: In each iteration of the  CA-ADMM algorithm \ref{Alg:cADMM_IRS}, the complexity of computing ${\bf x}^{k+1}$, ${\boldsymbol \lambda}_1^{k+1}$ and ${\boldsymbol \lambda}_2^{k+1}$ is $\mathcal{O}(N)$; the complexity of computing ${\bf y}_2^{k+1}$ is $\mathcal{O}(4N^3)$; according to \eqref{eqn:admmy}$\sim$\eqref{eqn:uyk}, the complexity of computing ${\bf y}_1^{k+1}$ is $\mathcal{O}(\max\{N_{\text B}N_{\text E}N^2,4N^3\})$ . Therefore, the overall complexity for computing ${\bf \Theta}$ in CA-ADMM algorithm is $\mathcal{O}(T_1\max\{N_{\text B}N_{\text E}N^2,4N^3\})$, where $T_1$ is the iteration number in Algorithm \ref{Alg:cADMM_IRS}.

 \textbf{Complexity of BCD-OMP Algorithms}: The complexity of BCD-OMP Algorithms is dominated by two aspects: 1) computing the FDB $\tilde{\bf W}^{\star}$, 2) the complexity of solving problem \eqref{eqn:hpproblem}. For the first aspect, the complexity of computing the FDB $\tilde{\bf W}^{\star}$ is $\mathcal{O}(T_2\max\{2N_{\text A}^3,2N_{\text A}^2N_{\text E}\})$ in the BCD-OMP algorithm, where $T_2$ is the iteration number of the BCD algorithm. For the second aspect, the BCD-OMP algorithm utilizes the low-complexity OMP algorithm for solving problem \eqref{eqn:hpproblem}, indicating that the complexity of second aspect is reasonable to be accepted.

\vspace{-10pt}
\section{Numerical Simulations}\label{Sec:simulation}
\vspace{-5pt}
In our simulations, channels ${\bf H}_{ij}$ is modeled as sparse channel \cite{Alkhateeb2014JSTSP}, where the number of paths is set as 4, the small scale propagation fading gain obeys `Rician' fading with $\kappa=13.2$.
The path loss exponents of the direct links from Alice to Bob and from Alice to Eve are 4, and the reflected links from Alice to IRS, from IRS to Bob and Eve are 2. The AoDs/AoAs and distances $d_{ij}$ are obtained from the locations of the Alice, Bob, Eve, and IRS, where $A=[0,5]$, $B=[60,0]$, $E=[45,0]$, $I=[55,5]$. The Bob's and Eve's noise power are $\sigma^2=-59$dBm.
Unless stated otherwise, we set the numbers of antennas and RF chains are $N_{\rm{A}}=N_{\rm{I}}=32$, $N_{\rm{B}}=2$, $N_{\rm{E}}=2$, $N_{\rm{RF}}=4$, $L_s=L_z=2$. The parameters $\alpha_B$, $L_y$, $\rho_1$, $\rho_2$, and $\epsilon_1$ in the CA-ADMM Algorithm \ref{Alg:cADMM_IRS} are set as $\alpha_B=\sigma^2$, $L_y=8$, $\rho_1=\rho_2=16$, and $\epsilon_1=10^{-5}$.

{\setlength\abovedisplayskip{1.5pt}
\setlength\belowdisplayskip{1.5pt}
  \begin{figure}\label{Fig:NIRS}
    \centering
  \centerline{\includegraphics[width=0.4\textwidth]{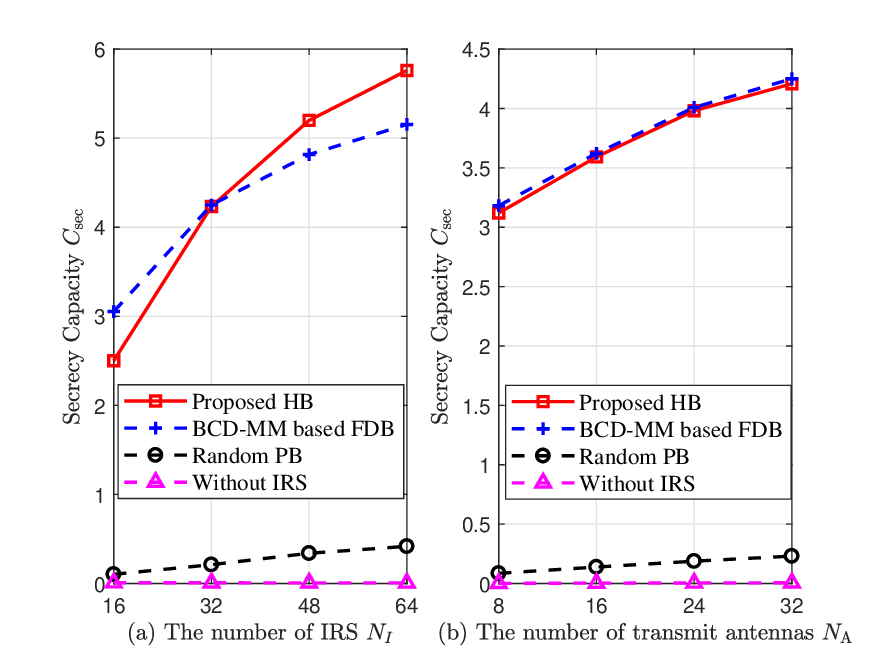}}
\caption{The SR comparisons among different algorithms versus (a) the number of IRS elements $N_{\rm I}$, (b) the number of antennas at Alice $N_{\rm{A}}$.}
\end{figure}}
Fig.1 depicts the SR of \emph{proposed CA-ADMM and BCD-OMP based HB strategy}, the \emph{BCD-MM based FDB strategy}, the \emph{random IRS strategy} and \emph{no IRS strategy}, versus the number of IRS elements $N_{\rm I}$ and the number of antennas $N_{\rm A}$, respectively.

As seen from the Fig.1(a), we have the following observations. First, when $N_{\rm I}=16$, the SC performance of the BCD-MM based FDB strategy is higher than the proposed HB strategy; while when $N_{\rm I}\geq32$, the proposed HB strategy outperforms the BCD-MM based FDB strategy. Moreover, the gaps of the SR performances achieved by the proposed HB strategy and the BCD-MM based FDB strategy increases as the increasing of $N_{\rm I}$. This observation demonstrates that the proposed HB strategy is more suitable for large-scale IRS scenarios. Second, the IRS assisted strategies outperform the no-IRS strategy. Thirdly, the performance gaps between the proposed HB strategy and random PB strategy are increasing as the increasing of the number of IRS elements $N_{\rm I}$.

As observed from Fig.1(b), under the varying number of antennas, the proposed HB strategy can achieve the similar performance to the BCD-MM based FDB strategy. Moreover, the gaps between the the proposed HB strategy and the random PB/without IRS strategies are increasing as the increasing of the number of antennas $N_{\rm A}$.

{\setlength\abovedisplayskip{1.5pt}
\setlength\belowdisplayskip{1.5pt}
\begin{figure}[htb]
  \centering
  \label{Fig:conver}
  \centerline{\includegraphics[width=0.35\textwidth]{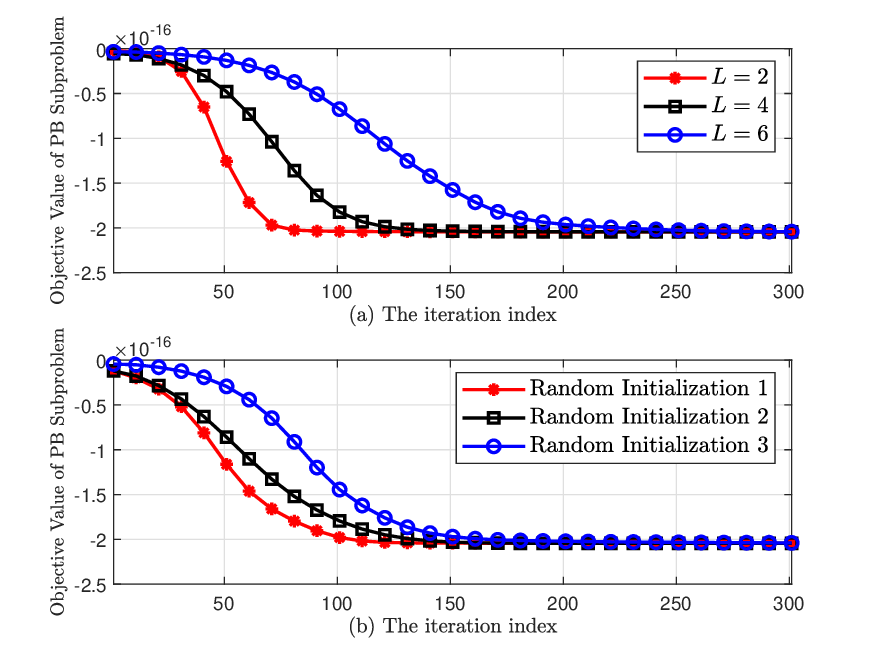}}
\caption{The convergence performance of the proposed CA-ADMM algorithm, where the parameters in each sub-figure are set as: (a) the initialization of all curves are the same, and $\rho_1=\rho_2=7L$; (b) $L=4$ and $\rho_1=\rho_2=28$.}
\end{figure}} 

Fig.2 shows the convergence performance of the proposed CA-ADMM algorithm. As observed from this figure, with the same initialization, the proposed CA-ADMM algorithm converges with different parameters $L$, $\rho_1$, and $\rho_2$ to the same objective value of the OF-PB subproblem. Moreover, although the initialization has impact on the required number of iteration, the CA-ADMM algorithm always converges to the same SC even with random initialization.

\end{document}